\documentclass[11pt,a4paper]{article}
\usepackage{amsfonts}
\usepackage{amsmath}
\usepackage{graphicx}

\input{tcilatex}
\begin{document}

\title{ENERGY BANDS SPLITTING \\
IN THE KOHMOTO MODEL}
\author{ O. BORCHIN \\
Department of Theoretical Physics, Faculty of Physics \\
West University of Timisoara, Bul. Vasile Parvan nr.4, RO-300223 \\
Timisoara, Romania \vspace{1mm}\\
E-mail: ovidiuborchin@yahoo.com \\
}
\date{}
\maketitle

\begin{abstract}
Peierls gap is analyzed in case of a two-dimensional lattice under the
influence of a magnetic field, in a tight-binding approximation. By using a
non-analytic class of potentials, such as the Kohmoto potential in the
Harper model, splitting effect occurs in the energy bands. In the
metal-insulator transition point, the charge carriers become separated due
to their energy, releasing and expanding the Peierls gap. As a result, the
energy bands around the Fermi level become localized in case of the
electrons and delocalized coresponding to the holes, since their energy
become lowered. These facts are supported by numerical investigations.
\end{abstract}

\begin{center}
recursive energy bands

(Received \today)
\end{center}

{\centering}

\section{INTRODUCTION}

\qquad Studies regarding two-dimensional electron systems under the
simultaneous influences of a periodic potential and of an external
homogeneous magnetic field have received much interest in the last four
decades \cite{azbel, wannier, wilkins, harper1, last}. This kind of systems
are important in solid state studies, because the nowadays nanoscale
technologies are based on the band engineering concept \cite{davies,
ashkroft}\ and on the thermodynamic formalism \cite{datta}. In order to
build heterostructures with a precise distribution of the energy bands, we
have to understand the way in which they appear.

The energy bands structure of the carriers in magnetic fields was predicted
in 1964 by Azbel, while in 1969 Langbein \cite{langbein}\ analyzing the
Landau levels, discovered \textquotedblleft the butterfly
pattern\textquotedblright , in fact a new and profound connection between
the admissible energy states and the form and size of the latticed pierced
by the magnetic field. This phenomenon was analyzed numerically by
Hofstadter \cite{hofstadter} in 1976. He obtained the self-similar structure
of the energy bands, with a shape of a butterfly.

It has been found that electrons moving under the simultaneous influence of
a periodic potential and of a magnetic field display a series of mini-bands,
separated by corresponding mini-gaps. The energy spectrum due to its unique
shape, represent even today a remarkable achievement of the theoretical
investigation in physics.

The interpretation given by Hofstadter to the numerical results, highlighted
the fact that for a given value of the anisothropy parameter, the system is
at a crossroad between the conductive and insulator states. This self-dual
point between localized and extended states, have been described by Andr\'{e}
and Aubry \cite{andre} too. In this critical point appears a gap described
by Peierls in 1955 \cite{peierl}\ known today as the Peierls gap \cite{ono,
kohat}. This represents ever since a cornerstone to any studies connected to
transitions between conductive and insulator states \cite{mahito}.

The aim of this paper is to investigate numerically the energy bands
distribution of a system in which the Peierls gap create a separation
between the charge carriers, leading from conductive to insulator states.
For this purpose we intend to analyze the Harper model with the Kohmoto
quasi-periodic potential. Similar studies have been done by Hiramoto and
Kohmoto \cite{hiramo}\ by resorting to a generalized Harper model.

The paper is organized as follows: After the Introduction, in Section II we
present as a short review of potentials of a two-dimensional lattice. In the
following one shows the Harper equation with the Kohmoto quasi-periodic
potential. The numerical results are discussed in Section III. In the last
Section we present the Conclusions of our research.

\section{Model description and numerical results}

\qquad In order to analyze the Harper equation with the Kohmoto
quasi-periodic potential, we have to reiterate the concept of potentials.
That's important because we are interested to identify the way in which
carriers cross the environment from a point to the other, depending on their
conductive properties.

Let us start from a one-dimensional system in a tight-binding approximation,
submitted to an external and homogeneous magnetic field \cite{weisz, papp}.
It can be viewed as a conductive media of a free electron gas. By
generalizing the above chain into a two-dimensional square lattice, we can
describe any particular direction using the Miller indices [$u$ $v$ $w$] 
\cite{tilley}. In this way, three kind of periodic potentials arise.

The first kind concerns the periodic potentials \cite{courta}\ which
corresponds to the directions [$0$ $1$ $0$] parallel to the $x$-axis, [$1$ $%
0 $ $0$] parallel to the $y$-axis and [$1$ $1$ $0$] for the first and second
bisectrices. These directions work in conjurrection with the $NN$ (nearest
neighbor) and the $NNN$ (next nearest neighbor) hoppings. The periodic
potentials are described in terms of the Bloch functions. They have a
rational number as slope, but not localized states. The second kind concerns
the super-periodic potentials \cite{pletico}, which occur to large $u$ of
the corresponding Miller indices [$u$ $1$ $0$]. These potentials have a
large integer as slope and lead to localized states.

The third one, the quasi-periodic potentials \cite{borgono, bellisard} have
a regular but never exactly repeating pattern with an irrational number as
slope. In this later case, the translation invariance is lost. One deal
instead with local rotational symmetries. It is a kind of long range order
potentials, with similarities between tiny bits appearing in different
places but slightly modified, from one to the other. The quasi-periodic
potentials can be found for instance,\ between [$u_{1}$ $1$ $0$] and [$u_{2}$
$1$ $0$] directions.

At the beginning these kind of potentials where only strange entities
without any applicability. Howewer, between 1962 and 1984 they became the
major description tool of a sort of quasi-crystals \cite{luck}. One of the
most intriguing properties of these solid state compounds are connected with
their behavior at low temperatures. It is known that any conductive medium
responds positively to the influence of low temperatures \cite{hyun} by
increasing its conductivity and giving birth to the superconductor state.
The quasi-periodic crystals behave vice-versa, by transforming from a
conductor state into an insulator one. Such a crystal is provided by
synthetic organic materials like $TTF-TCNQ$ (Tetra Thia Fulvalene - Tetra
CyaNoQuinodimethane) \cite{anderson}, which have been discovered between
1962 and 1973.

Besides the organic compounds, there are other kind of crystals which
exhibit a forbidden five-fold symmetry, such as the Al-Pd-Mn icosahedral
quasi-crystals \cite{shech}, discovered in 1984. This alloy, formed with 3
metals, each of them with good conductivity, is in fact almost an insulator.

An experimental attempt to understand the way in which electrons flow
through a quasi-periodic crystal have been done by Torres et al. \cite%
{torres} in 2003, resorting to a pulse propagation with a shallow fluid
covering a quasi-periodically drilled bottom. In order to implement the
quasi-periodicity, we have to resort to the Harper equation \cite{harper},
given by%
\begin{equation}
\psi _{n+1}+\psi _{n-1}+V_{n}\psi _{n+1}=E\psi _{n}
\end{equation}%
where $\psi _{n}$ is the wave functions at site $n$. Here $V_{n}$ is a
periodic function with the role of potential 
\begin{equation}
V_{n}=2\Delta \cos 2\pi (n\beta +\frac{k_{2}a}{2\pi })\text{ ,}
\end{equation}%
$\Delta $ being the anisothropy parameter and $E$ the allowed energies.

In other words we deal with a two-dimensional system under the influence of
an external and perpendicular magnetic field. The quotient $\beta =P/Q=\phi
/\phi _{0}$ is the commensurability parameter expressing the number of flux
quanta per unit cell, $P$ and $Q$ are two mutual prime integers and $k_{2}a$
stands for the Brillouin phase. Rational values of the commensurability
parameter, result in a finite number of energy bands. When it become an
irrational number, the energy bands becomes infinite, the spectrum depending
completely on the potential strength played by the anisothropy parameter.

So, for $\Delta >1$ we have localized solutions, for $\Delta <1$ we have
extended wave functions of a contiguous spectrum and for $\Delta =1$ we have
a singular continuous spectrum.

The Harper equation which originates from the minimal substitution on the
energy dispersion law, can be rewritten in terms of transfer matrices \cite%
{kohmoto1}\ as%
\begin{equation}
\left( 
\begin{array}{c}
\psi _{n+1} \\ 
\psi _{n}%
\end{array}%
\right) =T_{n}\left( 
\begin{array}{c}
\psi _{n} \\ 
\psi _{n-1}%
\end{array}%
\right)
\end{equation}%
where the transfer matrix \cite{negi}\ is given by%
\begin{equation}
T_{n}=\left( 
\begin{array}{cc}
E-V_{n} & -1 \\ 
1 & 0%
\end{array}%
\right)
\end{equation}%
while the potential is periodic $V_{n}(x)=V_{n}(x+1)$.

\section{Numerical studies}

\qquad In order to investigate the energy bands when the system passes from
a conductive state to an insulator one, we chose a particular class of
non-analytic potentials, namely the Kohmoto potential \cite{kadanof},
proposed independently by Ostlund \cite{ostlund}, because it preserves
discontinuities and takes only two values 
\begin{equation}
V_{n}\equiv \alpha =\QATOPD\{ . {V_{A}=1\text{ \ for }-\gamma <x\leq -\gamma
^{3}}{V_{B}=-1\text{ \ for }-\gamma ^{3}<x\leq \gamma ^{2}}
\end{equation}%
where $\gamma =(\sqrt{5}-1)/2$ represents the golden mean ratio. The Kohmoto
model was introduced at the beginning as a simple toy model \cite{hatsugai1,
satou}, with the purpose to explain the metal-insulator transition. By
considering only two kind of possible values of $V_{n}$, the Fibonacci
sequence can readily established. Accordingly, the potential becomes \cite%
{naumis} 
\begin{equation}
V_{n}=V_{B}+V_{A}(\left\lfloor (n+1)\beta \right\rfloor -\left\lfloor n\beta
\right\rfloor )
\end{equation}%
where $\left\lfloor n\beta \right\rfloor $ represents the greatest integer
lower than $n\beta $. Following the Naumis and Rodrigues work and using the
identity $n\beta =\left\lfloor n\beta \right\rfloor +\{n\beta \}$ the
potential can be rewritten as%
\begin{equation}
V_{n}=V_{A}\beta +V_{B}(1-\beta )+(V_{A}-V_{B})(\{n\beta \}-\{(n+1)\beta \})
\label{eq4}
\end{equation}%
where $\{n\beta \}$ denotes the decimal part.

In our numerical calculation the quantity $(V_{A}-V_{B})$ in (\ref{eq4})
known as the strength of the quasi-periodicity is taken as $(V_{A}-V_{B})=-1$%
. The next step is to introduce this potential into the Harper equation,
such as 
\begin{equation}
T_{n}=\left( 
\begin{array}{cc}
E-\alpha \beta -\alpha (1-\beta )+(\{n\beta \}-\{(n+1)\beta \}) & -1 \\ 
1 & 0%
\end{array}%
\right)
\end{equation}%
because the last one has many convenient symmetries, which can facilitate
the numerical analysis. In this way the metal-insulator transition can be
studied on the two-dimensional lattice, as a splitting effect of the energy
bands, corresponding to the electrons and holes concerning energy bands. The
numerical results are presented in $Fig.1$ for several values of $n$ as
follows:

$a)$ The energy bands are continuous. The conductive medium exhibits mixed
states of both electrons and holes if $n=1$.

$b)$ One deals with splittings in the energy spectrum, supporting the cosine
shape of the dispersion law of the energy if $n=1.1$.

$c)$ The energy gap tends to increase releasing the Peierls gap if $n=1.2$.

$d)$ More and more splittings occur in the inner bands distribution
releasing incrementally the Peierls gap. The system is still a conductive
medium if $n=1.5$.

$e)$ One deals with the Aubry duality point, which corresponds to the
transition state from conduction to the insulator one. The Fermi level is
located at the middle of the Peierls gap if $n=2$.

$f)$ Beyond the critical point the energy gaps tend to increase. By now the
system become more and more an insulator for $n=3$.

$g)$ The lower energy bands tend to decrease in an accentuated chaotic
manner, while the top bands tend to gather together around the zero value if 
$n=4$.

$h)$ The separations between the energy bands become more pronounced, while
the lower energy bands become a cloudy like configuration of points and
short range lines if $n=7$.

$i)$ Separations between energy bands are clear. Since an environment can't
be a conductive medium with only one carriers, determine the insulator
states. In this case state, the Peierls gap become to large between the
energy of the negative carriers and the positive ones, to allow a
combination. The conductive environment remain only with one kind of cherge
carriers, which leads to the\ suppression of current flow if $n=100$.%
\begin{equation*}
\FRAME{itbpF}{4.8518in}{4.9729in}{0in}{}{}{fig.1.eps}{\special{language
"Scientific Word";type "GRAPHIC";maintain-aspect-ratio TRUE;display
"USEDEF";valid_file "F";width 4.8518in;height 4.9729in;depth
0in;original-width 6.7942in;original-height 6.964in;cropleft "0";croptop
"1";cropright "1";cropbottom "0";filename '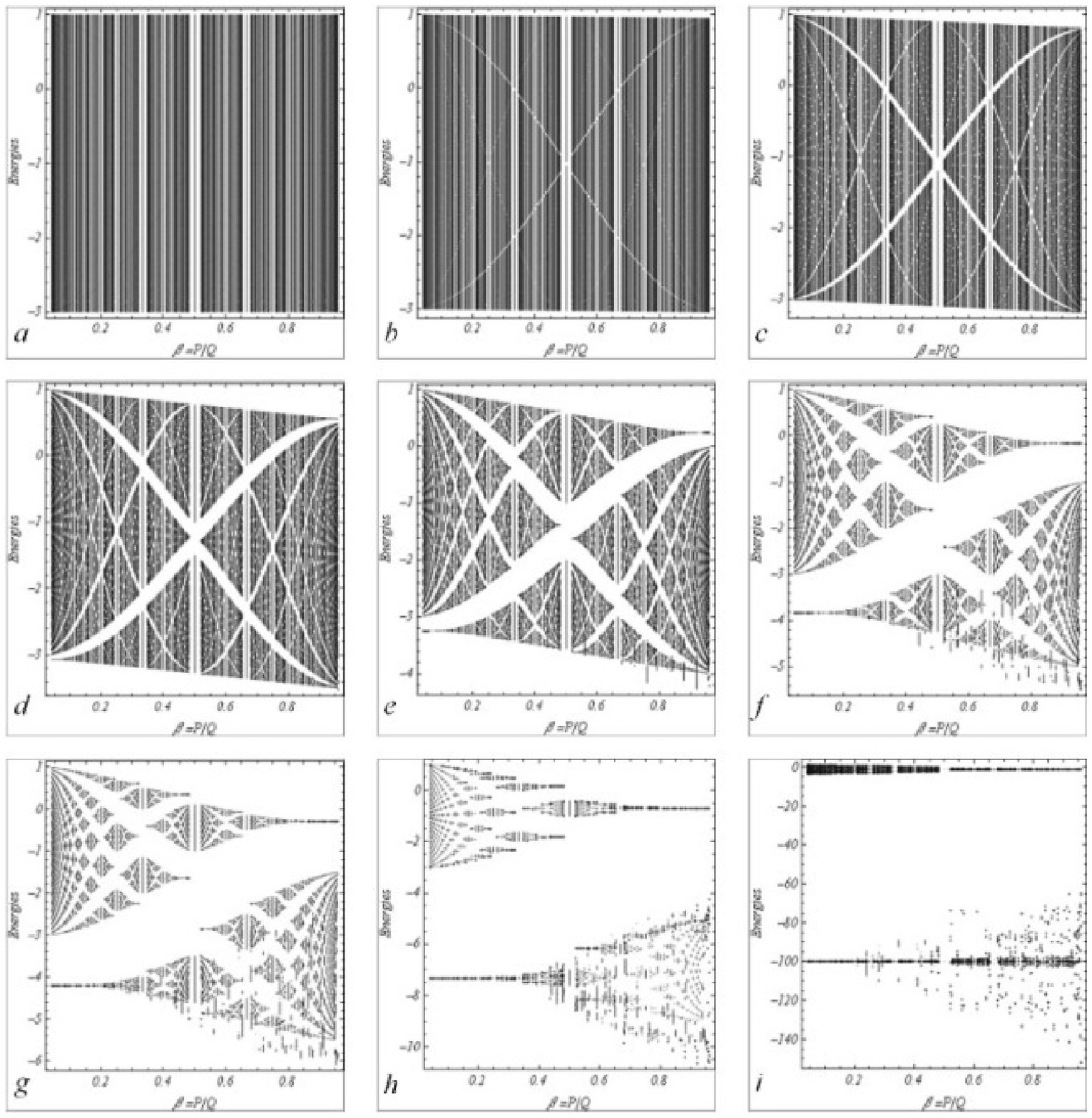';file-properties
"XNPEU";}}
\end{equation*}

\begin{center}
$Fig.1$ The energy spectrum versus $\beta $\ showing

the band splittings for several $n$-values.

$_{{}}$
\end{center}

By virtue of this transition the system exhibits less symmetric states, but
with a lower-energy configuration. The energy bands splits, due to the
periodic potential, which induce a small gap by releasing and expanding the
Peierls distortion. The band splittings take place in such a way that the
Fermi energy is always placed in the middle of the gap. The higher energy
bands corresponding to the electrons get gathered araund the zero energy,
while the negative ones corresponding to the holes lower their energy. This
interpretation of the energy band splittings is an atempt to investigate
numerically the way in which systems undergo transitions from conductive to
insulator states.

\section{Conclusions}

\qquad In this paper we analyzed numerically the way in which the energy
bands get splitting, when the system goes from conductive to the insulator
state, i.e. to a less symmetric but lower-energy configuration, passing
through the transition point. The energy bands become divided by increasing
small gaps in such a way that the Fermi energy will be placed always in the
middle of the gap. As result of $n$\ continuous variation, the system passes
from a symmetric state to symmetry broken ones. The higher energy bands
become locked, while the lower energy bands tend to become unlocked.

\section{Acknowledgment}

\qquad I wish to express my gratitude to Professor Erhardt Papp, for his
support.


\begin{thebibliography}{99}
\bibitem{azbel} M. Y. Azbel, Zh. Eksp. Teor. Fiz. 46, (1964) 939.

\bibitem{wannier} G. H. Wannier, Phys. Rev. 117 (1960) 432; Rev. Mod. Phys.
34 (1962) 645.

\bibitem{wilkins} M. Wilkinson, Proc. Roy. Soc. London A403, (1986) 153.

\bibitem{harper1} P.G. Harper, Proc. Roy. Phys. Soc. London A 68 (1955) 874.

\bibitem{last} Y. Last and M Wilkinson, J. Phys. A: Math. Gen 25 (1992) 6123.

\bibitem{davies} John H. Davies, \textsl{The physics of low-dimensional
semiconductors}, (Cambridge University Press, 1998).

\bibitem{ashkroft} N. W. Ashcroft, N.D. Mermin, \textsl{Solid State Physics}%
, (Saunders College, Phyladelphia, 1976).

\bibitem{datta} S. Datta, A. Sharma and R. Ramaswamy, Phys. Rev. E 68 (2003)
036104.

\bibitem{langbein} D. Langbein, Phys. Rev. 180 (1969) 633.

\bibitem{hofstadter} D.R. Hofstadter, Phys. Rev. B 14 (1976) 2239.

\bibitem{andre} G. Andr\'{e} and S. Aubry, Ann Isr. Phys. Soc. 3 (1980) 133.

\bibitem{peierl} R. Peierls, \textsl{Quantum theory of solids}, (Oxford
University Press, Oxford, 1955).

\bibitem{ono} Y. Ono and T. Hamano, J. Phys. Soc. Jpn. 69 (2000) 1769.

\bibitem{kohat} M. Kohmoto and Y. Hatsugai, Phys. Rev. B 41 (1990) 9527.

\bibitem{mahito} M. Kohmoto, Phys. Rev. Lett. 51/13 (1983) 1198.

\bibitem{hiramo} H. Hiramoto and M. Kohmoto, Phys Rev B Condens Matter.
40(12) (1989) 8225.

\bibitem{weisz} J. F. Weisz, Phys. Rev. B 44, (1991) 6515.

\bibitem{papp} E. Papp and. C. Micu, \textsl{Low-Dimensional Nanoscale
Systems on Discrete Spaces}, World Scientific, Singapore, 2007.

\bibitem{tilley} R. J. D. Tilley, \textsl{Crystals and crystal structures},
(John Wiley, New York, 2006).

\bibitem{courta} E. Courtade, O. Houde, Jean-Fran\c{c}ois Cl\'{e}ment, P.
Verkerk and D. Hennequin, Phys. Rev. A 74, (2006) 031403.

\bibitem{pletico} I. Pletikosi\'{c}, M. Kralj, P. Pervan, R. Brako, J.
Coraux, A. T. N'Diaye, C. Busse and T. Michely, Phys. Rev. Lett. 102, (2009)
056808.

\bibitem{borgono} F. Borgonovi and D. L. Shepelyansky, Physica D: Nonlinear
Phenomena 109/1-2 (1997) 24.

\bibitem{bellisard} J. Bellissard, B. Iochum and D. Testard, Commun. Math.
Phys. 141 (1991) 353.

\bibitem{luck} J. M. Luck and D. Petritis, J. of Statist. Phys. 42, 3/4
(1986) 289.

\bibitem{hyun} Hyun-Tak Kim, Kwang-Yong Kang, Bong-Jun Kim, Y. C. Kim, W.
Schmidbauer and J. W. Hodby, Physica C: Superconductivity 341, (2000) 729.

\bibitem{anderson} P. W. Anderson, P. A. Lee and M. Saitoh, Solid State
Communications 13 (1973) 595.

\bibitem{shech} D. Shechtman, I. Blech, D. Gratias, J.W. Cahn, Phys. Rev.
Lett. 53 (1984) 1951.

\bibitem{torres} M. Torres, J.P. Adrados, J.L. Arag\'{o}n, P. Cobo and S.
Tehuacanero, Phys. Rev. Lett. 90 (2003) 114501.

\bibitem{harper} P. G. Harper, Proc. Phys. Soc. London, Sect A 68 (1955) 879.

\bibitem{kohmoto1} M. Kohmoto, B. Sutherland and C. Tang, Phys. Rev. B 35
(1987) 1020.

\bibitem{negi} S. S. Negi and R. Ramaswamy, arXiv:nlin/0104054v1 [nlin.CD]
(2001).

\bibitem{kadanof} L.P. Kadanoff, M. Kohmoto and C. Tang, Phys. Rev. Lett. 50
(1983) 1870.

\bibitem{ostlund} S. Ostlund, R. Pandit, D. Rand, H.J. Schellhuber and E.D.
Siggia, Phys. Rev. Lett. 50 (1983) 1873.

\bibitem{hatsugai1} Y. Hatsugai and M. Kohmoto, J. Phys. Jpn. 61 (1992) 256.

\bibitem{satou} A. Satou and M. Yamanaka, Phys. Rev. B 63-21 (2001) 212403.

\bibitem{naumis} G. G. Naumis and F. J. L\'{o}pez-Rodriguez, Physica B
(2008) 1755.
\end{thebibliography}
\end{document}